\documentclass[a4paper]{jpconf}

\usepackage{amsmath,amssymb,mathrsfs,latexsym}
\usepackage{graphicx,epstopdf} 

\newcommand{\nn}{\nonumber}
\newcommand{\up}{\uparrow}

\newcommand{\bs}[1]{\boldsymbol{#1}}

\newcommand{\ket}[1]{\left|#1\right\rangle}
\newcommand{\bra}[1]{\left\langle#1\right|}

\def\b{\text{b}}
\def\r{\text{r}}
\def\g{\text{g}}
\def\y{\text{y}}
\def\c{\text{c}}
\def\m{\text{m}}

\newcommand{\ii}{\text{i}\,}
\def\ie{\emph{i.e.},\ }

\def\ea{\emph{et al.}}

\begin{document}
%---------------------------------------------------------------------
%\title{Spontaneous Breakdown of Parity in a Quantum Spin Chain}
\title{Spontaneous Parity Violation in a Quantum Spin Chain}
%---------------------------------------------------------------------

\author{Stephan Rachel$^1$, Dirk Schuricht$^2$, Burkhard Scharfenberger$^1$,  
 Ronny Thomale$^1$, and Martin Greiter$^1$}
\address{$^1$\,Institut f\"ur Theorie der Kondensierten Materie, 
Universit\"at Karlsruhe, 76128 Karlsruhe, Germany}
\address{$^2$\,Institut f\"ur Theoretische Physik A, RWTH Aachen, 
52074 Aachen, Germany}

%\ead{rachel@tkm.uni-karlsruhe.de}

\begin{abstract}
  We report on a spontaneous breakdown of parity in the ground state
  of a spin Hamiltonian involving nearest-neighbor interactions.   This
  occurs for a one-dimensional model where spins transform under the
  gauge field representation of QCD, the eight-dimensional adjoint
  representation of SU(3).  The ground state spontaneously violates
  parity and is two-fold degenerate.  In addition, the model possesses
  a non-vanishing topological string order parameter which we
  explicate analytically.
\end{abstract}

\section{Introduction}
Symmetries and conservation laws constitute an essential ingredient of
physical theories~\cite{Coleman85}.  An intriguing situation arises
when a given microscopic model Hamiltonian is known to be symmetric
under a certain type of transformation, but the expected regularity
associated with this symmetry does not manifest itself in the physical
properties.  Beyond anomalous breaking, the only other mechanism known
for this type of event is spontaneous symmetry breaking.  There, the
ground state of a system does not share the symmetries of the
Hamiltonian.  For continuous symmetries, a prominent example is the
ferromagnetic phase in the Heisenberg model, where the ground state
breaks the SO(3) rotational symmetry by singling out one certain
magnetization axis.

In this article, we consider the discrete parity symmetry
$\mathcal{P}$.  Parity symmetry breaking (PSB) for the weak
interaction has been a milestone in particle
theory~\cite{lee-56pr254}.  In the %condensed matter
theory of magnetism, PSB has emerged in different contexts for higher
dimensions, and is generally referred to the real space component of
chiral symmetry breaking~\cite{wen-89prb11413}, which additionally
includes time reversal symmetry breaking $\mathcal{T}$.  However, it
has remained elusive in most cases to define a model which
spontaneously generates PSB.  In the spin chain model which we discuss
here, the algebraic structure of the adjoint representation of SU(3)
at each site of the spin chain allows to accomplish this.  In this
contribution, we analyze the degenerate ground state manifold of
this model and study some low-energy properties.  Furthermore, we
observe that the model possesses a topological string order which we
calculate exactly.

\section{The model and its properties}
To construct a valence bond state with spins transforming under the
eight-dimensional adjoint representation $\bs{8}$ of SU(3), we place a
fundamental representation $\bs{3}$ and an anti-fundamental
representation $\bar{\bs{3}}$ of SU(3) on each lattice site.  We then
project the resulting tensor product onto the symmetric subspace,
which yields the adjoint representation:
$\mathcal{S}\{\bs{3}\otimes\bar{\bs{3}}\}=\bs{8}$.  We generate an
overall spin singlet state by coupling each representation $\bs{3}$
antisymmetrically with a representation $\bar{\bs{3}}$ on the
neighboring site into a singlet bond:
$\mathcal{A}\{\bs{3}\otimes\bar{\bs{3}}\}=\bs{1}$.  This construction
yields two linearly independent representation $\bs{8}$ states,
$\Psi^\text{L}$ and $\Psi^\text{R}$, which may be visualized as
\begin{equation}\nn
\setlength{\unitlength}{1pt}
\begin{picture}(220,30)(-20,-10)
\put(-45,7){$\ket{\Psi^\text{L}}\equiv$}
\multiput(5,10)(25,0){9}{\circle{4}}
\multiput(12,10)(25,0){9}{\circle{6}}
\multiput(9,10)(25,0){9}{\circle{15}}
\thicklines
\multiput(15,10)(25,0){8}{\line(1,0){13}}
\put(-3,10){\line(1,0){6}}
\put(215,10){\line(1,0){6}}
\thinlines
\put(58.5,1){\line(0,-1){9}}
\put(58.5,-13){\makebox(1,1){\small one site}}
\end{picture}
\end{equation}
and its parity conjugate obtained by interchanging fundamental (small
circles) and anti-fundamental representations (larger circles).  The
big circles indicate a lattice site and the horizontal lines between
the sites are singlet bonds.  As the construction is analogous to AKLT
states~\cite{aklt}, we may write the state vectors $\Psi^\text{L}$ and
$\Psi^\text{R}$ as matrix product states.  Taking (\b,\r,\g ) and
(\y,\c,\m ) as bases for the representations $\bs{3}$ and
$\bs{\bar{3}}$, respectively, we obtain the matrix~\cite{greiter-07prb184441}
\begin{equation}\nn
%  \label{eq:m8explictly}
  M_i%^{\bs{8}}
  =\left(\!\begin{array}{ccc} 
  \frac{2}{3}\ket{\b\y}_i-\frac{1}{3}\ket{\r\c}_i-\frac{1}{3}\ket{\g\m}_i
  &\ket{\r\y}_i&\ket{\g\y}_i\\[4pt]
  \ket{\b\c}_i
  &-\frac{1}{3}\ket{\b\y}_i+\frac{2}{3}\ket{\r\c}_i-\frac{1}{3}\ket{\g\m}_i
  &\ket{\g\c}_i\\[4pt]
  \ket{\b\m}_i&\ket{\r\m}_i
  &-\frac{1}{3}\ket{\b\y}_i-\frac{1}{3}\ket{\r\c}_i+\frac{2}{3}\ket{\g\m}_i 
  \\[4pt]    
    \end{array}\!\right).
\end{equation}
Assuming periodic boundary conditions (PBCs), the representation
$\bs{8}$ states $\Psi^\text{L}$ and $\Psi^\text{R}$ are hence given by
the trace of the matrix products
\begin{equation}
  \label{eq:8VBSL}
  \ket{\Psi^\text{L}}
  =\text{tr}\biggl( \prod_i M_i%^{\bs{8}}
  \biggr) \qquad {\rm and} \qquad
  \ket{\Psi^\text{R}}
  =\text{tr}\biggl( \prod_i M_i%^{\bs{8}}
  ^\text{T} \biggr).
\end{equation}
These states transform into each other under space reflection
or parity symmetry: $\mathcal{P}\Psi^\text{L}=\Psi^\text{R}$, 
%and $\mathcal{P}\Psi^\text{R}=\Psi^\text{L}$, 
where the discrete parity
transformation is defined as $\mathcal{P}:\bs{S}_i \rightarrow \bs{S}_{N-i+1}$.
%\begin{equation}
%\mathcal{P}:\bs{S}_i \rightarrow \bs{S}_{N-i+1}.
%\end{equation}
Here $\bs{S}_i$ is an eight-component spin operator corresponding to
the eight-dimensional representation of SU(3) and $N$ denotes the number of
lattice sites.   
Eigenstates of the parity operator with eigenvalues $\pm 1$
are given by the even (symmetric) and odd (antisymmetric) superpositions 
%\begin{equation}
$\Psi^{\pm}=\frac{1}{\sqrt{2}}\left( \Psi^\text{L} \pm \Psi^\text{R}\right)$.
%\end{equation}

\begin{figure}[t!]
  \centering
  \includegraphics[scale=0.63]{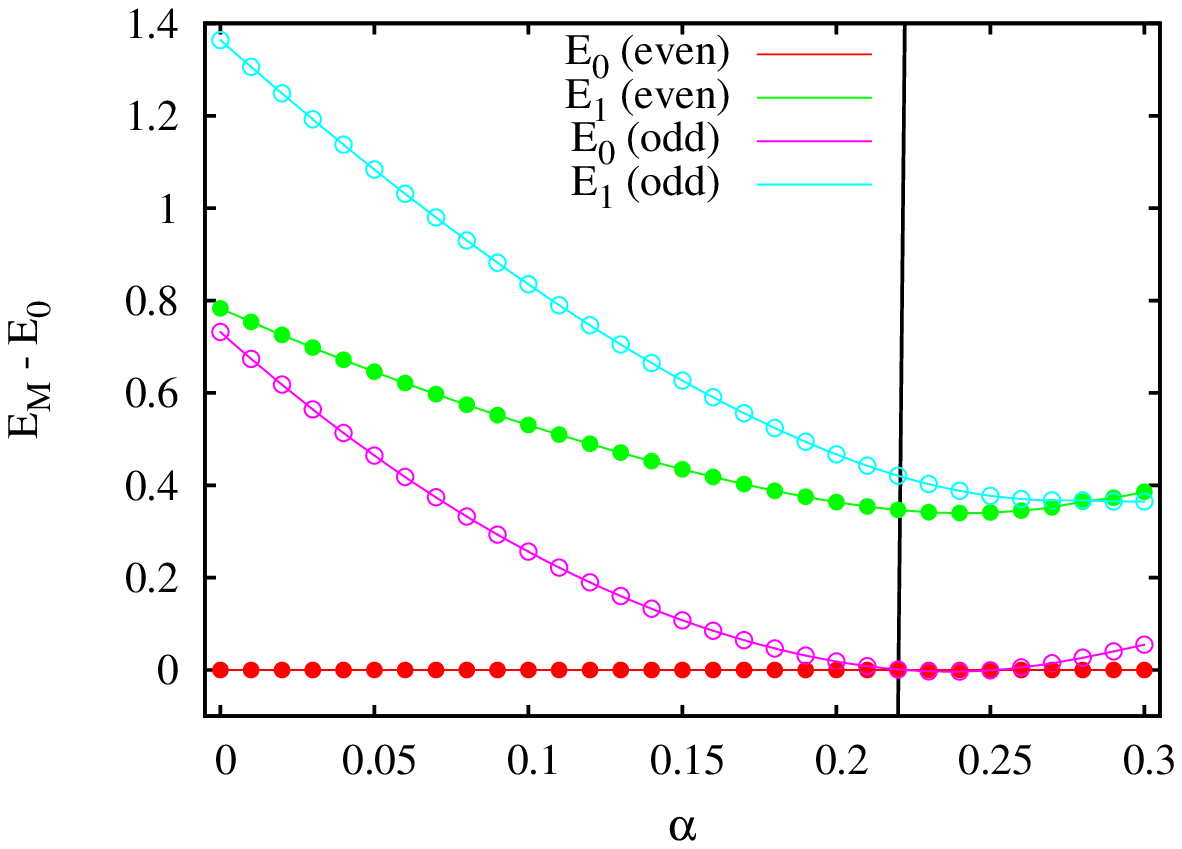}~
  \includegraphics[scale=0.63]{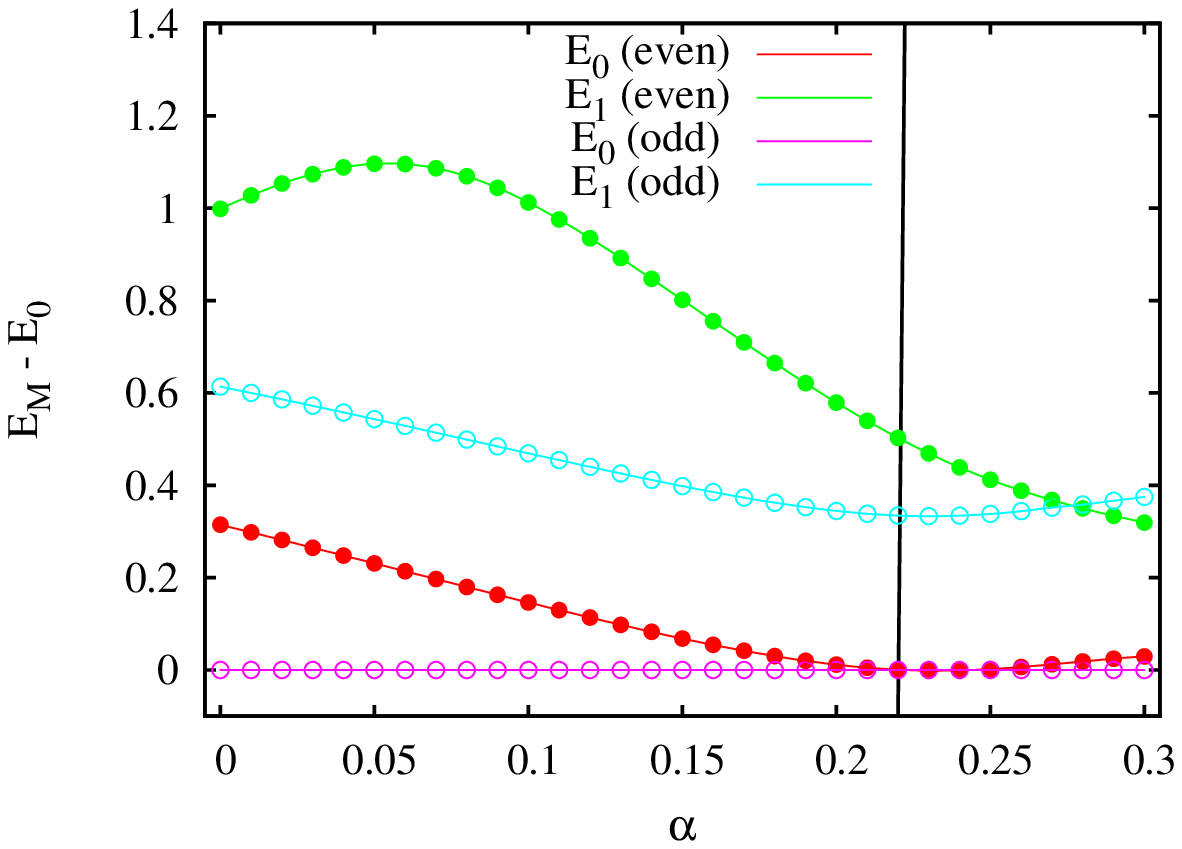}\\[-10pt]
\setlength{\unitlength}{5pt}
\begin{picture}(0,0)(0,0)
\put(-7.8,28){\makebox(0,0){$N=8$}}
\put(38.8,28){\makebox(0,0){$N=9$}}
\put(-12.3,3.5){\makebox(0,0){$\up$}}
\put(-12.4,0.6){\makebox(0,0){$\frac{2}{9}$}}
\put(34.3,3.5){\makebox(0,0){$\up$}}
\put(34.4,0.6){\makebox(0,0){$\frac{2}{9}$}}
%\put(-20.8,7){\makebox(0,0){$\Psi^-$}}
\end{picture}
\caption{(Color online) Spectral flow of the two lowest energies of
  the two parity sectors (even/odd) under variation of the prefactor
  $\alpha$ of the biquadratic Heisenberg term in
  $\mathcal{H}(\alpha)=\sum_i \bs{S}_i\bs{S}_{i+1} + \alpha
  (\bs{S}_i\bs{S}_{i+1})^2$.  PBCs have been imposed for both even
  ($N=8$, left plot) and odd ($N=9$, right plot) number of sites.  The
  spectrum of the Heisenberg model differs considerably for even and
  odd number of sites.  In the region around the exact model
  \eqref{ham.8VBS}, $\alpha=2/9$, the spectra become similar.  As we
  move from \eqref{ham.8VBS} to the Heisenberg point, the ground-state
  degeneracy is lifted.  For an even (odd) number of sites $\Psi^+$
  ($\Psi^-$) becomes the ground state for $0\le\alpha< 2/9$.}
  \label{fig:flow}
\end{figure}

A parent Hamiltonian which annihilates the states $\Psi^{\pm}$ is given
by~\cite{greiter-07prb060401}
\begin{equation}
\label{ham.8VBS}
\mathcal{H}=\sum_{i=1}^N\left(\,\bs{S}_i\bs{S}_{i+1}\,+
\frac{2}{9}\bigl(\bs{S}_i\bs{S}_{i+1}\bigr)^2 + 1 \right).
\end{equation}
The parity operator $\mathcal{P}$ commutes with the Hamiltonian,
$[\mathcal{H},\mathcal{P}]=0$, while the ground states
$\Psi^\text{L/R}$ or $\Psi^\pm$ spontaneously violate this symmetry.
The model \eqref{ham.8VBS} and its entropy were studied by Katsura
\ea~for open boundary conditions (OBC) and fixed dangling
spins\,\cite{Korepin}, while similar models with other symmetries were
introduced in Refs.\,\cite{arovas08prb,tu1,DR}.  Starting from
\eqref{ham.8VBS}, it is natural to ask what happens for a
representation $\bs{8}$ Heisenberg model with pure bilinear
interactions $\bs{S}_i\bs{S}_{i+1}$.  We find numerically that the
ground state degeneracy is lifted when the prefactor of the
biquadratic Heisenberg interaction is decreased from $2/9$ to 0 (see
Fig.\,\ref{fig:flow}).
Preliminary finite size scaling data indicate that the splitting
between the two lowest lying states of the Heisenberg model vanishes
as we approach the thermodynamic limit $N\to\infty$, as required for a
spontaneous symmetry violation.  Above the two degenerate ground
states, we expect a Haldane-type gap for the lowest lying
excitations~\cite{greiter-07prb184441}.  This conclusion is further
supported by finite size scaling data not included here as well as the
exponential decay of the static spin-spin correlations evaluated in
the following section.

\section{String order parameter}

In analogy to Refs.~\cite{DR,denNijsRommelse89,tu2}, we define the
string operators
\begin{equation}
O_{1j}^{ab}=-J_1^a\,\exp\Biggl[\ii\pi\sum_{k=2}^{j-1}
\Bigl(J_k^3+\frac{2}{\sqrt{3}}J_k^8\Bigr)\Biggr]\,J_j^b,
\quad a,b\in\bigl\{3,8\bigr\}.
\end{equation}
With the matrix product representations \eqref{eq:8VBSL}, the string
order can be calculated using the method introduced by
Kl\"umper~\emph{et al.}~\cite{Kluemper-91} for the $q$-deformed model.
The analysis proceeds in several steps.  First, we calculate the norm
of the matrix product state $\Psi^\mathrm{L}$.  We introduce the
complex conjugated matrix $\tilde{M}$ according to
$\tilde{M}_{\sigma\sigma'}=M^*_{\sigma\sigma'}$, \ie by taking the
complex conjugate of each matrix element of $M_i$ without transposing
the matrix $M_i$.  We then define the $9\times 9$ transfer matrix $R$
at any lattice site as
\begin{equation}
\label{eq:Rmatrix}
R_{\alpha\beta}=R_{(\sigma\tau),(\sigma'\tau')}=
\tilde{M}_{\sigma\sigma'}\,M_{\tau\tau'},
\end{equation}
where we order the indices as $\alpha,\beta=1,\ldots,9 \leftrightarrow
(11),(12),\ldots,(33)$.  Finally, the norm is given by
\begin{equation}
\label{eq:VBSnorm}
\bra{\Psi^{\text{L}}}\Psi^{\text{L}}\rangle=
\text{tr}\Bigl(R^N\Bigr)=\frac{1}{3^N}\Bigl(8^N+7\,(-1)^N\Bigr)
\rightarrow\left(\frac{8}{3}\right)^N,
\end{equation}
where have evaluated the trace by diagonalization of $R$.  Second, we introduce
the transfer-matrix representation of the spin operators $J^{3,8}$ by
\begin{equation}
\hat{J}^{3,8}_{\alpha\beta}=\hat{J}^{3,8}_{(\sigma\tau),(\sigma'\tau')}=
\tilde{M}_{\sigma\sigma'}\,J^{3,8}\,M_{\tau\tau'}.
\end{equation}
Third, we introduce the operator
$A=\tilde{M}\,e^{\ii\pi(J^3+2J^8/\sqrt{3})}\,M$ such that
\begin{equation}
  \big\langle O_{1j}^{33}\big\rangle_\mathrm{L}\equiv
  \frac{\bra{\Psi^{\text{L}}}O_{1j}^{33}\ket{\Psi^{\text{L}}}}
  {\bra{\Psi^{\text{L}}}\Psi^{\text{L}}\rangle}=
  -\frac{\text{tr}\Bigl(\hat{J}^3A^{j-2}\hat{J}^3R^{N-j}\Bigr)}
  {\bra{\Psi^{\text{L}}}\Psi^{\text{L}}\rangle}
  \stackrel{N\rightarrow\infty}{\longrightarrow} 
  \frac{9}{64}\left(1+(-1)^j\frac{32}{8^j}\right)
  \stackrel{j\rightarrow\infty}{\longrightarrow}\frac{9}{64}.
\end{equation}
In the same way we obtain in the limit $N\rightarrow\infty$
\begin{equation}
  \big\langle O_{1j}^{88}\big\rangle_\mathrm{L}=\frac{3}{64}
  \left(1+(-1)^j\frac{224}{8^j}\right),\quad
  \big\langle O_{1j}^{38}\big\rangle_\mathrm{L}=
  \big\langle O_{1j}^{83}\big\rangle_\mathrm{L}^*=
  -\text{i}\frac{3\sqrt{3}}{64}\left(1-8\frac{(-1)^j}{8^j}\right)
\end{equation}
as well as $\big\langle O_{1j}^{ab}\big\rangle_\mathrm{R}= \big\langle
O_{1j}^{ba}\big\rangle_\mathrm{L}$.  Note that the expectation values
of the string order parameters remain finite in the limit
$j\rightarrow\infty$.  In analogy to the original AKLT
model~\cite{KennedyTasaki92prb}, we attribute this string order as
well as the 18-fold degeneracy of the ground state of a chain with
open boundary conditions to the violation of a discrete symmetry.
Obviously, by choosing $A=R$, one obtains the static correlation
functions $\langle J_1^aJ_j^b \rangle=
-\delta_{ab}\,(-1)^j\,\frac{27}{2}\,8^{-j}$, \ie we find exponentially
decaying correlations with correlation length $\xi=1/\ln 8$.

%\newpage
\section{Generalization to SU($n$)}
The model \eqref{ham.8VBS} is a special case of a model of SU($n$) spins 
transforming under the $n^2-1$ dimensional adjoint representation 
$\mathcal{S}\{\bs{n}\otimes\bar{\bs{n}}\}$ %=\bs{n^2-1}$ 
with Hamiltonian
\begin{equation}
\mathcal{H}_{\text{SU}(n)} = \sum_i \left( \bs{S}_i\bs{S}_{i+1} +
\frac{2}{3n}\left( \bs{S}_i\bs{S}_{i+1} \right)^2 + \frac{n}{3} \right).
\label{eq:SUn}
\end{equation}
The ground state is again two-fold degenerate and exhibits exponentially 
decaying correlations with %correlation length 
$\xi=1/\ln{(n^2-1)}$.  
Note that $n^2-1$ is just the dimension of the adjoint representation.
% Note that the prefactor of the biquadratic term vanishes
% for $n\to\infty$ and the parity breaking model \eqref{eq:SUn} reduces
% to the simple Heisenberg model.  This supports our conjecture
% that the pure Heisenberg model has a two-fold degenerate ground state
% due to the broken parity symmetry.
We conjecture that the corresponding Heisenberg model has a two-fold
degenerate ground state due to the broken parity symmetry as well.

\section{Conclusion}
In this contribution, we have established an example of spontaneous
parity violation in a quantum spin chain.  We further identified a
non-local string order parameter for the model considered.

\section*{Acknowledgments}
This work was supported in part by the DFG Forschergruppen 912 (DS) and 
960 (BS and MG).

\section*{References}
%\bibliographystyle{/users/tkm/rachel/bib/prsty}
%\bibliography{/users/tkm/rachel/bib/paper,/users/tkm/rachel/bib/book,/users/tkm/rachel/bib/htc,/users/tkm/rachel/bib/martin,/users/tkm/rachel/bib/na}

\begin{thebibliography}{9}

\bibitem{Coleman85}
S. Coleman, {\em Aspects of symmetry} (Cambridge University Press, Cambridge,
  1985).

\bibitem{lee-56pr254}
T.~D. Lee and C.~N. Yang, Phys. Rev. {\bf 104},  254  (1956); \emph{ibid.}
  \textbf{106}, E1371 (1957).

\bibitem{wen-89prb11413}
X.~G. Wen, F. Wilczek, and A. Zee, Phys. Rev. B {\bf 39},  11413  (1989).

\bibitem{aklt} 
I. Affleck, T. Kennedy, E.~H. Lieb, and H. Tasaki, 
Phys. Rev. Lett.  {\bf 59}, 799 (1987);
Commun. Math. Phys. {\bf 115},  477  (1988).

\bibitem{greiter-07prb184441}
M. Greiter and S. Rachel, Phys. Rev. B {\bf 75},  184441  (2007).

\bibitem{greiter-07prb060401}
M. Greiter, S. Rachel, and D. Schuricht, Phys.~Rev.~B {\bf 75},  060401(R)
  (2007).

\bibitem{Korepin}
H. Katsura, T. Hirano, and V.~E. Korepin, J. Phys. A: Math. Theor. {\bf 41},
135304 (2008).

\bibitem{arovas08prb}
D.~P. Arovas, Phys. Rev. B {\bf 77}, 104404 (2008).

\bibitem{tu1}
H.-H. Tu, G.-M. Zhang, and T. Xiang, Phys. Rev. B {\bf 78}, 094404 (2008).

\bibitem{DR}
D. Schuricht and S. Rachel, Phys. Rev. B {\bf 78}, 014430 (2008).

\bibitem{denNijsRommelse89}
M. den Nijs and K. Rommelse, Phys. Rev.~B {\bf 40},  4709  (1989);
S.~M. Girvin and D.~P. Arovas, Physica Scripta {\bf T27},  156  (1989).

\bibitem{tu2}
H.-H. Tu, G.-M. Zhang, and T. Xiang, preprint {\tt arXiv:0807.3143} (2008).

\bibitem{Kluemper-91}
A. Kl\"umper, A. Schadschneider, and J. Zittartz, 
J.~Phys.~A: Math. Gen. {\bf  24},  L955  (1991);
Z.~Phys.~B {\bf 87},  281  (1992).

\bibitem{KennedyTasaki92prb}
T. Kennedy and H. Tasaki, Phys. Rev.~B {\bf 45},  304  (1992);
Commun. Math. Phys. {\bf 147},  431  (1992);
M. Oshikawa, J.~Phys.: Condens. Matter {\bf 4},  7469  (1992).

\end{thebibliography}

\end{document}